\documentclass[twocolumn]{svjour3}          
\usepackage{graphicx}
\usepackage{amsmath,amssymb}
\usepackage{color}
\journalname{Quantum Information with Neutral Particles}
\begin{document}

\title{Multibit C$_k$NOT quantum gates via Rydberg blockade
}

\titlerunning{Multibit C$_k$NOT quantum gates via Rydberg blockade}        

\author{L. Isenhower$^1$, M. Saffman$^1$, and K. M\o lmer$^2$ }

\authorrunning{L. Isenhower, M. Saffman, and K. M\o lmer} 

\institute{1) Department of Physics, University of Wisconsin,
1150 University Avenue, Madison, Wisconsin 53706, USA,  
2) Lundbeck Foundation Theoretical Center for Quantum System Research,
Department of Physics and Astronomy, University of Aarhus, DK-8000 \AA rhus C, Denmark       
}

\date{Received: date / Accepted: date}

\maketitle

\begin{abstract}
Long range Rydberg blockade interactions have the potential for efficient implementation of quantum gates between multiple atoms. Here we present and analyze a protocol for  implementation of a $k$-atom controlled NOT (C$_k$NOT) neutral atom gate. This gate can be implemented using sequential or simultaneous addressing of the control atoms which requires only $2k+3$ or $5$ Rydberg $\pi$ pulses respectively.   A detailed error analysis relevant for
implementations based on alkali atom Rydberg states is provided
which shows that gate errors less than 10\% are possible for $k=35$.  
\end{abstract}

\section{Introduction}
\label{intro}

Quantum gates and entanglement of pairs of neutral Rb atoms have been demonstrated in the last few years using Rydberg blockade interactions\cite{Gaetan2009,Urban2009,Wilk2010,Isenhower2010,Zhang2010}. The idea of using transient excitation of Rydberg states to mediate  strong interactions was introduced more than 10 years ago\cite{Jaksch2000}
and has attracted a strong interest due to the potential for implementation of a wide variety of quantum computing and quantum communication protocols\cite{Saffman2010}. 

The fundamental Rydberg blockade two-qubit gate relies on the fact that excitation of a single atom to a Rydberg level can block the subsequent excitation of a second atom when the dipolar interaction between two Rydberg atoms is much stronger than the light-atom coupling driving the excitation. If the Rydberg excitation lasers  are resonant for transitions between one of the qubit levels and the Rydberg level, the excitation of the first atom is dependent on the qubit state, and in this way the excitation of the second atom also becomes conditional on the state of the first atom. Conditional evolution of the second atom is sufficient for creating two-atom entanglement and, together with single qubit operations, provides a library of gates that can be used for universal quantum computation. 

Although any desired unitary evolution of a multi-bit register can be implemented by combining sequences of the one- and two-qubit universal gates\cite{Barenco1995} such a procedure is often inefficient. For example the Toffoli gate or C$_2$NOT gate where  the logical AND of two control atoms determines the transformation of the target atom state appears  frequently in quantum algorithms. Unfortunately a single  Toffoli gate requires at least six CNOT gates  for its implementation\cite{Nielsen2000} which increases the circuit complexity. 

Rydberg blockade interactions have the very useful property that excitation of a single control atom can simultaneously block the excitation of many target atoms, provided they all are close enough, that is within a so-called ``blockade radius"\cite{Lukin2001}, from the control atom. 
This radius can be on the order of tens of microns which could reach many qubits in both 2D and 3D optical dipole trap arrays.
The blockade interaction has been utilized to devise a number of protocols for multi-bit encoding schemes\cite{Brion2007d} and quantum gates\cite{Moller2008,Muller2009,Saffman2009b,Weimer2010}  that provide direct routes to entanglement of several qubits at a time.

In this contribution we show that the $k$ control atom (C$_k$NOT) generalization of the Toffoli gate can be implemented efficiently using Rydberg blockade interactions between multiple atoms. Our most efficient implementation requires only 5 Rydberg $\pi$ pulses, independent of $k$,  which can be contrasted with the best known circuit implementation of a C$_k$NOT which requires $32k-120$ elementary 1- and 2-qubit gates for $k\ge5$\cite{Maslov2003}. It has further been demonstrated that a C$_k$NOT requires at least $2k+2$ CNOT gates\cite{Shende2009} which shows that the 6 CNOT construction of the Toffoli gate in \cite{Nielsen2000} is optimal. We have explicitly shown that our C$_k$NOT protocol can be used for efficient implementation of Grover's quantum search\cite{Molmer2011}, and we anticipate that it will also be useful for other quantum algorithms.

In Sec. \ref{sec.gates} we describe two different implementations of a Rydberg C$_k$NOT gate using sequential and simultaneous addressing of the control atoms, and discuss the major sources of error during gate operation. In Sec. \ref{sec.errors} we analyze the sequential addressing approach in detail and give analytical and numerical estimates for the gate errors using parameters relevant for implementation with cold alkali atoms. The analysis is repeated for the simultaneous addressing approach in Sec. \ref{sec.errors2}. We conclude in Sec. \ref{sec.conclusion} with a discussion of the results obtained.

\section{C$_k$NOT pulse sequences}
\label{sec.gates}

\begin{figure}
\begin{centering}
\includegraphics[width=7.cm]{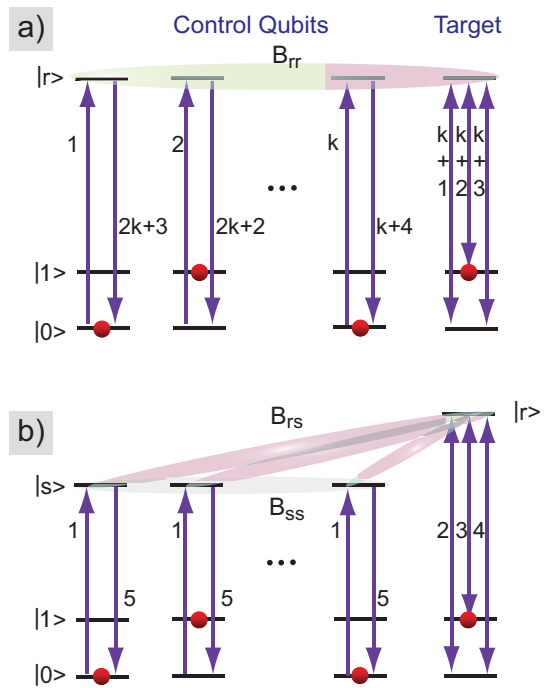}
\caption{\label{fig:gate}C$_k$NOT pulse sequences. a) Sequentially addressed C$_{k}$NOT gate where the numbers label the order of the pulses applied to the control bits. For this sequence the first control qubit in the $|0\rangle$ state will block all the remaining pulses until it gets de-excited. This blockade prevents the target state swap. Therefore the target state is swapped only when all the control qubits are in the $|1\rangle$ state. 
b) Simultaneously addressed version. Here all the control bits are excited to a Rydberg state $|s\rangle$ that does not interact strongly with the same Rydberg state allowing any control qubit in the $|0\rangle$ state to be excited. Next the target state is swapped via a Rydberg state $|r\rangle$ which has a strong interaction with the $|s\rangle$ state. Finally the control qubits are returned to the ground state. Therefore if any control qubit is in $|0\rangle$  the target state swap pulses are blockaded.}
\end{centering}
\end{figure}

The C$_{k}$NOT gate can be implemented using Rydberg blockade in two ways as shown in Figure \ref{fig:gate}. If we are limited to using individual addressing and a single Rydberg state, then the gate is implemented by applying a series of Rydberg $\pi$ pulses to the control qubits in order, i.e. from 1 to $k$, between the $|0\rangle$ state and an appropriate Rydberg state $|r\rangle$.   Then a sequence of three $\pi$ pulses to this Rydberg level is applied to the target qubit that swaps the target state as long as no Rydberg blockade is present, and finally the control qubits are returned to their original state with a phase shifted $\pi$ pulse to each control qubit in reverse order, i.e. $k\rightarrow 1$ as shown in Figure \ref{fig:gate}a. The total number of Rydberg $\pi$ pulses is thus $2k+3.$ This $2k+3$ pulse sequence was originally presented in \cite{Saffman2010} Sec. III.C.3.

If we have the ability to excite to two different Rydberg levels and can simultaneously address the control qubits then another possibility emerges. Here all of the control qubits are excited to a mutually non-interacting Rydberg level $|s\rangle$ followed by swapping the state of the target qubit via a Rydberg level $|r\rangle$ that is blockaded by any control qubit which happens to be in the $|s\rangle$ state. Finally the control qubits are returned to their initial state again using a phase shifted $\pi$ pulse.  This implementation requires only five $\pi$ pulses independent of $k$. As we will see this is advantageous since the amount of time spent in the Rydberg level is reduced at large $k$. However, it is difficult to find pairs of interacting and non-interacting states with sufficient asymmetry $({\sf B}_{rs}\gg {\sf B}_{ss})$ to obtain very high gate fidelities. For the Toffoli gate $(k=2)$ an alternative scheme using three different Rydberg levels is also possible, as was proposed in \cite{Brion2007c}.

For either approach we could replace the three pulses on the target atom with a single 
$2\pi$ pulse to implement a C$_k$Z gate which would require a total of $(2k+2)$ or four $ \pi  $ pulses, and then obtain a C$_k$NOT by adding Hadamard rotations to the target atom before and after the gate. Both approaches, using CZ or the controlled swap approach of Fig. \ref{fig:gate}, were demonstrated for two-atom gates in \cite{Isenhower2010}.

\section{Error estimates for sequential addressing}
\label{sec.errors}

In this section we develop detailed error estimates for the sequential addressing version of C$_k$NOT. When such a gate is used in a quantum computer the errors will vary depending on the 
state which the gate operates on. Some input states will lead to higher errors than others. Since the number of states is exponentially large an analysis which calculates the error separately for each possible state quickly becomes unwieldy. We therefore follow the simpler procedure of  averaging over all possible input states without regard to the spatial location of the information in the qubit array. The resulting error estimates should therefore be considered average errors: specific input states may have substantially higher errors. 

Furthermore we consider only the intrinsic errors due to the physical implementation of the gate. Additional technical noise sources such as laser pulse fluctuations, finite temperature Doppler and atomic spatial position effects, and magnetic fluctuations are not considered. These additional noise sources may be significant, particularly when using the gate to create entangled states\cite{Saffman2011}, but their analysis lies outside the scope of this contribution. 

\subsection{Intrinsic errors}

There are several intrinsic error sources for  implementation of the C$_{k}$NOT gates. These errors arise from the finite lifetime of the Rydberg states and undesired excitations due to insufficient blockade shift or off-resonant excitation from the other qubit level.  To derive the intrinsic errors we follow the methodology of \cite{Saffman2005a,Saffman2010} and  assume that the errors from each source are small and can be considered independently. Then each contribution can be added together to obtain a total error. Finally, to obtain the error for the gate, we average the errors over all  possible computational basis input states. 

First we consider the individual addressing case shown in Fig. \ref{fig:gate}a). We will calculate in turn the following errors: a) spontaneous emission from Rydberg states for control (c) and target (t) atoms, and b) pulse rotation errors from finite blockade shift or off-resonant excitation of the other qubit state.  For error a) each control qubit undergoes a $\pi$ pulse between the $|0\rangle$ and Rydberg states, a wait period, and a final $\pi$ pulse back to the ground state. For each $\pi$ pulse the spontaneous emission error is given by $\frac{\pi}{2\Omega\tau}$ where $\Omega$ is the Rabi frequency between the ground and Rydberg state and $\tau$ is the Rydberg state lifetime. During the waiting period the Rydberg state can decay giving an error of $\frac{t}{\tau}$ where $t$ is the time spent in the Rydberg state. In addition higher order terms occur when the qubit is excited by an off resonant pulse either because of finite blockade shifts or because the qubit was in the $|1\rangle$ state which is not connected to the Rydberg levels. The pulse scheme gives an error of $\frac{\pi}{2\Omega\tau}+\frac{n_{\rm wait}\pi}{\Omega\tau}+\frac{\pi}{2\Omega\tau}$ when the qubit is in the $|0\rangle$ state. In this formula $n_{\rm wait}$ is the number of $\pi$ pulses that happen between the two pulses that target the  control qubit $i$ which can be found to be $n_{wait}=3+2\left(k-i\right)$ where the 3 comes from the 3 $\pi$ pulses applied to the target qubit and the other term accounts for the pulses applied to the $i+1$ to $k$ control qubits. Next we must count the number of input basis states that have this error and average this over the computational basis states. This error can only occur when the  qubit $i$ is in  $|0\rangle$  and no earlier control qubits were in this state to cause Rydberg blockade. This number of states is found to be $2^{k+1-i}$ for  control bit $i$  and since there are $2^{k+1}$ possible computational states we find that by averaging over the input states we get a factor of $1/2^{i}$. Putting this together we find that the leading order error term due to spontaneous emission for the control qubits is 

$$
E_{se,c,1}=\frac{\pi}{\Omega\tau}\sum_{i=1}^{k}\frac{1}{2^{i}}\left\{1+\left[3+2\left(k-i\right)\right]\right\}=\frac{2\pi}{\Omega\tau}k 
\nonumber
$$

There is an additional spontaneous emission error from off resonant excitation due to blockade leakage. For this case we do not have full excitation to the Rydberg state and the amount of population is given by $\Omega^{2}/2{\sf B}^{2}$ where ${\sf B}$ is the blockade shift of the Rydberg state. For this we  assume that the blockade shift for all qubits can be given by some common minimum blockade shift and thus we will  conservatively calculate the maximum possible error. The amount of error is the same as the resonantly excited case just multiplied by the excitation probability of $\Omega^{2}/2{\sf B}^{2}$; however, the number of states seeing this error is different. Accounting for this difference we find 

\begin{eqnarray}
E_{se,c,2}&=&\frac{\pi\Omega}{2{\sf B}^{2}\tau}\sum_{i=2}^{k}\frac{1+3+2\left(k-i\right)}{2^{k+1}}\sum_{j=1}^{i-1}2^{k-j}\nonumber\\
&=&\frac{\pi\Omega}{4{\sf B}^{2}\tau}\left(k^{2}-k\right).
\label{eq.sec2}
\end{eqnarray}

The spontaneous emission errors for the target qubit are a little easier to find as counting the number of states for each error is easier. The target sees a pulse scheme with three Rydberg $\pi$ pulses. First a pulse from $|0\rangle$ to Rydberg, then from $|1\rangle$, and finally from $|0\rangle$ again. We assume that the Rabi frequency for both ground states is the same. For the lowest order term we find

$$
E_{se,t,1}=\frac{\pi}{\Omega\tau}\frac{1}{2^k}
$$
 and the blockade leakage term is found to be 
$$
E_{se,t,2}=\frac{5\pi\Omega}{4{\sf B}^{2}\tau}\frac{2^{k}-1}{2^{k+1}}.
$$

The other type of errors b) can be termed rotation errors. These are due to off-resonant excitations which cause residual undesired rotations between the ground and Rydberg states. The leading term is due to blockade leakage and is proportional to $\Omega^{2}/2{\sf B}^{2}$. Since the blockade shift is generally small compared to the qubit frequency for implementations involving alkali hyperfine states the term proportional to $\Omega^{2}/2\omega_{10}^{2}$ is smaller, where $\omega_{10}$ is the ground state splitting. Performing an analysis  similar to that for the  spontaneous emission 
errors we find

\begin{eqnarray}
 E_{r,c,1}&=&\frac{\Omega^{2}}{{\sf B}^{2}}\frac{1}{2^{k+1}}\sum_{i=1}^{k-1}2^{k}-2^{i}\nonumber\\
&=&\frac{\Omega^{2}}{2{\sf B}^{2}}\left(k-2+\frac{1}{2^{k-1}}\right)
\nonumber
\end{eqnarray}

\begin{eqnarray} E_{r,c,2}&=&\sum_{i=1}^{k}2^{-i}\frac{\Omega^{2}}{\omega_{10}^{2}}+\sum_{i=2}^{k}\frac{\Omega^{2}}{\left(\omega_{10}\pm {\sf B}\right)^{2}}\frac{1}{2^{k+2}}\sum_{j=0}^{i-2}2^{k-j}\nonumber\\
&=&\frac{\Omega^{2}}{\omega_{10}^{2}}\left( 1-\frac{1}{2^{k}}\right) +\frac{\Omega^{2}}{2\left(\omega_{10}\pm {\sf B}\right)^{2}}\left(k-2+\frac{1}{2^{k-1}}\right)\nonumber
\end{eqnarray}

\begin{eqnarray}
E_{r,t,1}&=&\frac{3\Omega^{2}}{2{\sf B}^{2}}\sum_{i=1}^{k}\frac{2^{-i}}{2}\nonumber\\
&=&\frac{3\Omega^{2}}{4{\sf B}^{2}}\left( 1-\frac{1}{2^{k}}\right)
\nonumber\\
E_{r,t,2}&=&\frac{1}{2^{k}}\frac{\Omega^{2}}{2\omega_{10}^{2}}+\frac{2^{k}-1}{2^{k}}\frac{3\Omega^{2}}{2\left(\omega_{10}\pm {\sf B}\right)^{2}}
\nonumber
\end{eqnarray}

 It is important to note that we have accounted for changes in the effective ground state splitting due to the blockade shift. In general the blockade shift for this off resonant state will be different from the blockade shift for the targeted ground state due to other nearby Rydberg levels but for simplicity we have assumed that it is still  given by the same minimum blockade shift.

Combining the eight errors listed above we get a total C$_k$NOT error of 
\begin{eqnarray}
E&=&\frac{\pi\Omega}{4{\sf B}^2\tau}\left[k^2-k+\frac{5}{2}\left(1-\frac{1}{2^{k}}\right) \right]
+\frac{ 2\pi  }{\Omega\tau}\left(k+\frac{1}{2^{k+1}}\right)\nonumber\\
&&
+\frac{ \Omega^2}{2 {\sf B}^2}\left(k -\frac{1}{2}+\frac{1}{2^{k+1}}\right)\nonumber\\
&&
+\frac{ \Omega^2}{2 ({\sf B}\pm \omega_{10})^2}\left(k+1-\frac{1}{2^k} \right)
 +\frac{\Omega^2}{\omega_{10}^2}\left(1-\frac{1}{2^{k+1}}\right) .\nonumber\\
\label{eq.Et1}
\end{eqnarray}
Several comments can be made about the total error. 
The Rydberg state spontaneous emission error proportional to $1/\Omega\tau$ enters with weight $k$ due to the time needed to attempt excitation of  all  $k$ control atoms, plus a term $\sim 1/2^k$ from the target atom which is the small probability that the target atom is not blockaded.  
Thus the target atom error spontaneous emission error is negligible. 
We see that all terms except the first one  scale linearly or slower with $k$. The error contribution which is  quadratic in  $k$
comes from (\ref{eq.sec2}) for the spontaneous emission of the control bits. There are $k$ control bits, and the amount of time an average bit spends in a Rydberg state is proportional to $k$, hence the error scales as $k^2.$ The prefactor multiplying this term is smaller by a factor of $1/(\Omega\tau)\ll 1$ than the second line in Eq. (\ref{eq.Et1}) which implies that for moderate values of $k$ the error grows linearly in $k$, after which it will switch over to quadratic in $k$ for large $k$. 

The gate error is minimized by choosing the optimum value of $\Omega$
with respect to the available blockade shift ${\sf B}$ and Rydberg lifetime $\tau.$
The relative scalings for implementation of Rydberg blockade gates are $\omega_{10}\gg {\sf B}\gg \Omega \gg 1/\tau.$ Dropping the terms $\sim 1/\omega_{10}^2$ and working in the limit of ${\sf B}\tau\gg k$ we find the optimum Rabi frequency to be 
\begin{equation}
\Omega_{\rm opt}\simeq(2\pi)^{1/3}\frac{{\sf B}^{2/3}}{\tau^{1/3}}
\label{eq.Omegaopt}
\end{equation}
  which  is the same scaling as for the CNOT gate\cite{Saffman2005a}. Numerical checks show that the optimum frequency found numerically agrees with $\Omega_{\rm opt}$ to within about 10\% provided ${\sf B}\tau>10 k.$

Using $\Omega_{\rm opt}$ we obtain an approximate error estimate of 
\begin{equation}
E_{\rm opt}\simeq \frac{3\pi^{2/3}}{2^{1/3}}\frac{k}{({\sf B}\tau)^{2/3}}+
\frac{\pi^{4/3}}{2^{8/3}}\frac{k^2}{({\sf B}\tau)^{4/3}}.
\label{eq.Eopt}
\end{equation}
The $({\sf B}\tau)^{-2/3}$ scaling is the same as that found for the CNOT gate\cite{Saffman2005a}. For  $({\sf B}\tau)^{2/3}\gg k$  the
C$_k$NOT gate error grows linearly in $k$. The less favorable quadratic in $k$ error scaling takes over when $k$ becomes comparable to $({\sf B}\tau)^{2/3}.$

Equations (\ref{eq.Et1},\ref{eq.Eopt}) are based on the assumption that 
all pairs of atoms have the same blockade interaction ${\sf B}$. 
In a real implementation in an optical trap array or lattice the 
pairwise separations will range from some minimum to maximum value, and it is therefore necessary to take the spatial arrangement of atoms into account. We provide this analysis in Sec. \ref{sec.lattice} below.

\begin{figure}
\begin{centering}
\includegraphics[width=.9\columnwidth]{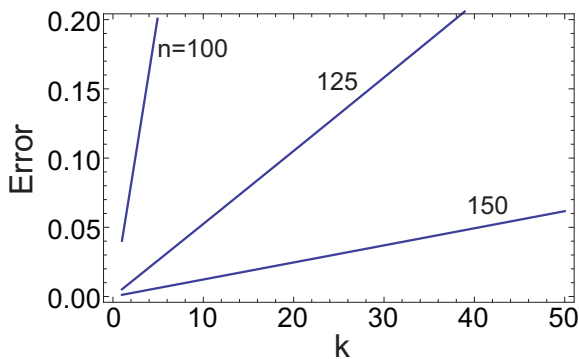}
\par\end{centering}

\caption{\label{fig.error} C$_k$NOT error assuming the same interaction strength for all atom pairs in  the sequential addressing scheme for Cs ns states with $n=100,125,150$. Parameters used were $\tau=(330,540,820)~\mu\rm s$, ${\sf B}/2\pi=(0.69,9.0,52.)~\rm MHz,$ and $\omega_{10}/2\pi=9200~\rm MHz.$ The blockade shift values were calculated for a fixed two-atom separation of $R=20~\mu\rm m$. With the above parameters  we find  $\Omega_{\rm opt}/2\pi=(0.11,0.53,1.5)~\rm MHz.$}
\end{figure}

Nevertheless, if we simply assume that all atom pairs have the same blockade shift as the weakest interacting pair we will obtain some insight into the order of magnitude of the gate errors.  
As an example of this type of estimate let us consider Cs ns Rydberg states in a lattice where the largest two-atom separation is $R=20~\mu\rm m$. The C$_k$NOT errors for $n=100,125,150$ are shown in Fig. \ref{fig.error}.
 The displayed curves were calculated using the approximate result (\ref{eq.Eopt}) as well as from (\ref{eq.Et1}) with $\Omega\rightarrow\Omega_{\rm opt}.$
On the scale of the figure there is no discernible difference between the two error estimates.  We see that the error decreases rapidly with 
principal quantum number and is about $0.06$ at $k=50$ for the 150s state. The error grows quadratically at much larger values of $k$ than shown in the figure. We emphasize that the errors shown in the figure are only meant to be thought of qualitatively since they assume all atom pairs have the  blockade shift of the pair with the largest separation. In order to find a more accurate error estimate we must take into account the physical placement of the qubits as is done in the next section.

\subsection{Lattice averaging}
\label{sec.lattice}

\begin{figure}
\begin{centering}
\includegraphics[width=.35\columnwidth]{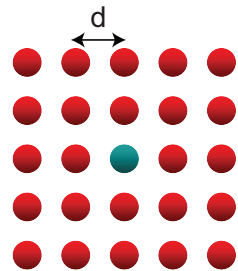}
\par\end{centering}

\caption{\label{fig.lattice}Two-dimensional qubit lattice with period $d$. The central atom  serves as the target qubit. }
\end{figure}

We will assume that the atoms are placed on a two-dimensional square lattice with period $d$, as shown in Fig. \ref{fig.lattice}. While a three dimensional system would give a larger number of atoms within a blockade sphere, it appears difficult to achieve  individual control of each atom  in such a geometry. We will therefore concentrate on a two-dimensional implementation.

There are several factors that limit the minimum value of $d$ that may be used. In order to avoid collisions between a Rydberg excited electron and a neighboring ground state atom we require $d> \sim a_0 n^2$ with $a_0$ the Bohr radius. A precise evaluation of the maximum number of atoms that can be placed within a blockade sphere while  meeting this requirement 
can be found in \cite{Saffman2008} where it was shown that in 2D $k'\sim n^{2/3}$ at constant gate error (we  use $k'=k+1$ for the number of atoms).

This scaling needs to be modified by the additional limitation  that ${\sf B}(d)$, the blockade shift between atoms separated by $d$, should satisfy  ${\sf B}(d)<[U(n)-U(n-1)]/2\sim n^{-3}$, where $U(n)$ is the energy of Rydberg level $n$. If ${\sf B}(d)$ exceeds this amount then 
the F\"orster resonance between pairs of atoms with $n$ differing
by $\pm 1$ may lead to an interaction induced resonance, which would break the blockade effect. 
Setting ${\sf B}(d_{\rm min})=[U(n)-U(n-1)]/2$ with resonant F\"orster scaling  
${\sf B}(d)\sim n^4/d^3$ gives $d_{\rm min}\sim n^{7/3}$. 
This is slightly more restrictive than the collisional limit in the preceding paragraph. 
When the largest two-atom separations are in the van der Waals regime the blockade scaling is ${\sf B}(R)\sim n^{12}/R^6$ for the heavy alkalis\cite{Saffman2008} which leads to an $R$ limit at constant gate error of $R_{\rm max}\sim n^{7/3}.$ The number of atoms  scales as $k'\sim (R_{\rm max}/d_{\rm min})^2\sim \rm constant.$ We  conclude that, asymptotically for large $n$, the number $k'$ is constant 
and there is no advantage to working with very large values of $n$. 
Indeed, the numerical results given below show that the optimum performance is obtained for $n\sim 75$. 

The reason for this preference for moderate values of $n$ is somewhat subtle and can be understood as follows. 
The F\"orster interaction ${\sf B}(n,n')$ is different between an atom pair $(n,n)$ and an atom pair $(n,n'\ne n)$. Calculations show that for $ns$ states with $n>100$ the interaction varies smoothly with $n'-n$ and
${\sf B}(n,n)>{\sf B}(n,n'\ne n)$ for $|n'-n|$ small. However, the difference in interaction strength for $n'=n-1$ is only about $20\%$ so the limit on $d_{\rm min}$ indicated above could only be slightly relaxed. This situation is qualitatively different at smaller $n$ since the Cs $ns$ F\"orster defects increase rapidly for $n<\sim 80.$\cite{Walker2008}. It turns out that the limit on $d_{\rm min}$ is then no longer set by the need to avoid interaction induced resonances, but by 
 limits on optical beam focusing, and the need to avoid crosstalk between neighboring sites. This limit is not fundamental and can be improved on by using beams with specific spatial profiles. For definiteness we will assume a lower limit of $d_{\rm min}=1~\mu\rm m$.
With a fixed $d_{\rm min}$ there are diminishing returns in going to  lower $n$ since the interaction gets too weak. We find numerically that the optimum tradeoff is in the vicinity of $n=75.$

\begin{figure}
\begin{centering}
\includegraphics[width=.9\columnwidth]{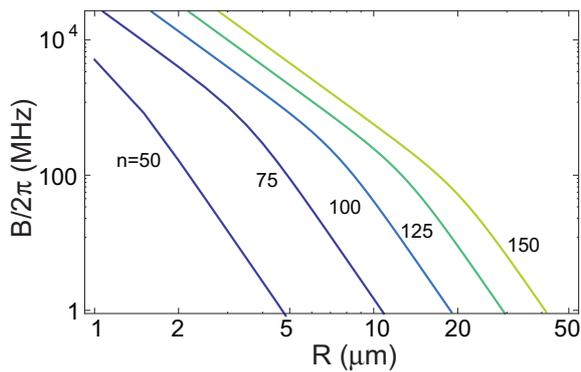}
\par\end{centering}
\caption{\label{fig.BvsR}Blockade shift for several Cs $ns$ levels. The blockade is calculated as described in \cite{Walker2008}.}
\end{figure}

We present representative numerical results for the gate error  for Cs $50s - 150s$ states in Table \ref{tab.Et1}. The errors are found  by averaging  Eq. (\ref{eq.Et1}) over all values of $R_{ij}=|{\bf R}_i-{\bf R}_j|$ that occur in a square lattice, using the ${\sf B}(R)$ curves from 
Fig. \ref{fig.BvsR}.  
This gives an error which depends on $\Omega$ which we then set to minimize the error. 
In this way we capture both the resonant dipole-dipole and van der Waals behavior.  For $n=50,75$ we set $d=1.~\mu\rm m$. For $n\ge 100$ we improve on the resonant interaction limit described above by reducing $d$ such that 
${\sf B}(d)=1.5\times [U(n)-U(n-1)]$. 
In a square lattice the next separation is $R=\sqrt2 d$ which gives a
blockade shift of ${\sf B}(d)/2^{3/2}\simeq 0.5 \times [U(n)-U(n-1)]$.
Further optimization with ${\sf B}(d)$ several times larger than the gap may be possible but this 
requires additional fine tuning of the architectural parameters which is a task we leave for future 
work\cite{RydbergBadLuck}.

\begin{table}[!t]
\centering
\begin{tabular}{|c|c|c|c|c|c|c|c|c|}
\hline 
 $n$&$\tau~(\mu\rm s)$&$d~(\mu\rm m)$&$k=3$& $8$&$15$&$24$&$35$\\
\hline
50 & 63.& 1. &.003&.04&.30&$>1$&$>1$\\
75 & 170.&1. &.0003&.0017&.0071&.026&.078\\
100 & 330.&2.13 &.0009&.0021&.010&.035&.11\\
125 & 540.&3.65& .0003 &.0024&.011&.039&.12\\
150 & 820.&5.66&.0004 &.0028&.012&.042&.13\\
\hline
\end{tabular}
\caption{Sequential addressing C$_k$NOT gate errors from Eq. (\ref{eq.Et1}) averaged over a square lattice for several different Rydberg Cs $ns$ levels. For $n=75$ the Rabi frequency was set to $\Omega/2\pi = (45,29,20,15,12) ~\rm MHz$ for $k=(3,8,15,24,35).$  }
\label{tab.Et1}
\end{table}

\begin{figure}
\begin{centering}
\includegraphics[width=.9\columnwidth]{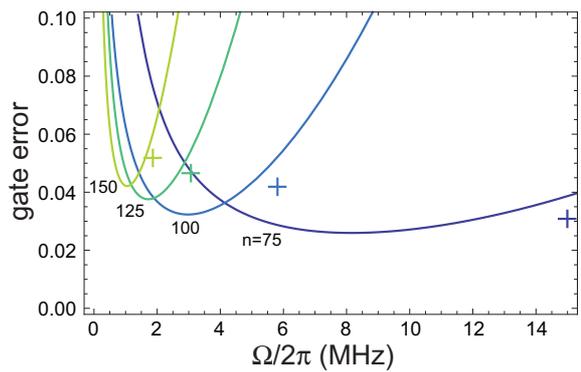}
\par\end{centering}
\caption{\label{fig.Evsomega}Dependence of the gate error on $\Omega$ for $k=24$ and several Cs $ns$ levels. The crosses near each curve mark the values of $\Omega_{\rm opt}$ and $E_{\rm opt}$ found from lattice averaging of Eqs. (\ref{eq.Omegaopt},\ref{eq.Eopt}).}
\end{figure}

We see that for $k\le8$ the error is approximately constant with $n$ whereas for $k\ge 15$ the $75s$ level gives the best results. 
We obtain an intrinsic  gate error less than 0.1 for $k=35$.
The variation of the error with $\Omega$ is shown in Fig. \ref{fig.Evsomega} for $k=24.$ At lower $n$ the Rydberg lifetime is shorter and the optimum Rabi frequency is about 8 MHz, decreasing down to about 1 MHz at $n=150$. The numerically found optima are about 20\% lower than predicted by Eq. (\ref{eq.Eopt}) and they occur at about twice smaller Rabi frequencies than predicted by Eq. (\ref{eq.Omegaopt}).
The difference between the analytical estimates and numerics can be attributed to lattice averaging of the interaction in the numerical solution. The error minima are sufficiently broad that no unreasonable fine tuning of $\Omega$ is needed.

We emphasize that the estimated gate errors represent  averages over states of the $k+1$ bits in the computational basis. Specific input states will have gate errors that are larger than the average by some factor. In order to assess the strength of our results it is desirable to put some bound on the error variations.  We first point out that our error estimates are conservative since we essentially compute the probability that ``something goes wrong", and assign zero fidelity to the resulting state (even if it may actually be an algorithmically useful  outcome despite the error).

To be more quantitative we can estimate the error for what is plausibly a worst case input state. Let us suppose a single control bit is Rydberg excited, but that it appears first in the Rydberg excitation sequence, and thereby spends the longest possible time in the Rydberg state, and that it has the largest possible spatial separation from the target bit.    To estimate the error for such a case we use the parameters from Table \ref{tab.Et1}, $n=75$   and use Eq. (26) from \cite{Saffman2010}  for the gate error.   We calculate the blockade shift $\sf B$ from Fig \ref{fig.BvsR} for the largest two atom separation that occurs for  $k=35$ and artificially decrease the Rydberg lifetime $\tau$ by putting $\tau \rightarrow \tau/35$ to account for the longer time spent in the Rydberg state. 
Doing all this gives a C$_2$NOT error of 0.02 which is still less than the average error in Table \ref{tab.Et1}. 
This may indicate that our estimate of the average  
error is, indeed, a conservative one. An identification of what are  
the worst possible input states, and what is the best choice of  
parameters to reduce the dominant errors remains an open question for  
future work.

If we were to implement the C$_{35}$NOT gate with elementary gates we would need at least 72 CNOT gates\cite{Shende2009}. 
It is known from random benchmarking studies\cite{Knill2007} that the total error  grows approximately linearly  with the number of gates, so an equivalent error for the C$_{35}$NOT gate would require CNOT gates each with an error of $0.001$ which is 
well beyond the current state of the art for gate error which has been achieved in ion traps\cite{Benhelm2008b}. 

As can be seen from Table \ref{tab.Et1} the gate errors grow faster than linearly in $k$ which differs from the result of 
Eq. (\ref{eq.Eopt}). There are two reasons for this discrepancy. 
First the ratio of $k/({\sf B}\tau)^{2/3}$ is not very small for the largest $k$ values considered so there is some quadratic scaling. More significantly, since there is a very large variation of $\sf B$ across the array 
it is not possible to simultaneously optimize all errors with a single $\Omega$ value, which violates the assumption leading to 
(\ref{eq.Eopt}).

\section{Error estimates for simultaneous  addressing}

\label{sec.errors2}

The simultaneous addressing scheme of Fig \ref{fig:gate}b) can be analyzed along similar lines. Parameters associated with the control and target qubits are labeled with subscripts c and t, respectively.  We identify the following five leading error sources due to spontaneous emission and pulse rotation errors on the control and target qubits:
\begin{eqnarray}
E_{se,c}&=&\frac{\pi}{2\Omega_c\tau_c}k+\frac{3\pi}{2\Omega_t\tau_c}k,\label{eq.Esim1}\\
E_{se,t}&=&\frac{\pi}{\Omega_t\tau_t}\frac{1}{2^k},\\
E_{r,c,1}&=&\frac{{\sf D}_{cc}^2}{\Omega_c^2}\frac{k}{2^{k+1}}
\sum_{j=1}^{k-1}\begin{pmatrix}k-1\\j\end{pmatrix}j^2\nonumber\\
&=&\frac{{\sf D}_{cc}^2}{16\Omega_c^2}(k^3-k),\\
E_{r,c,2}&=&\frac{{\Omega}_{c}^2}{2\omega_{10}^2}k,\\
E_{r,t}&=& \frac{3{\Omega}_{t}^2}{4{\sf B}_{ct}^2}\frac{1}{2^k}\sum_{j=1}^k \begin{pmatrix}k\\j\end{pmatrix}\frac{1}{j^2}\nonumber\\
&+& \frac{3{\Omega}_{t}^2}{4{\sf B}_{ct}^2}\frac{1}{2^k}\sum_{j=1}^k \begin{pmatrix}k\\j\end{pmatrix}\frac{1}{(j+\omega_{10}/{\sf B}_{ct})^2}. \label{eq.Esim5}
\end{eqnarray}
The binomial sums that appear in several terms account for averaging over multiple Rydberg excited control atoms.  
The blockade shift between control and target atoms is ${\sf B}_{ct}$  and ${\sf D}_{cc}$ is the direct interaction between two control atoms which we calculate as the dipole-dipole interaction strength between the laser excited states. Parameters will be chosen to ensure that \\
$|{\sf D}_{cc}/\Omega_{cc}|< 1$ so there is no blockade between control atoms. 
We assume the blockade and interaction shifts are additive in the number of Rydberg excited control atoms, which can be as large as $k$. 
For $k$ large the issue of interaction induced Rydberg resonances which were discussed in the previous section may again be problematic, as well as nonlinear behavior of multiple interacting Rydberg atoms\cite{Pohl2009}. While  detailed study of such effects is outside the scope of this work, we will see that the simultaneous addressing version of the gate can be operated in a regime where ${\sf B}_{ct}(d), {\sf D}_{cc}(d)$ are small compared to the Rydberg gap $U(n)-U(n-1)$ which adds confidence to the validity of our present results.  
Note that $\Omega_c$ is only constrained by the off-resonant excitation limit of $\Omega_c^2/\omega_{10}^2$ so we expect that the minimum error will be found at large values of $\Omega_c.$  


\begin{figure}
\begin{centering}
\includegraphics[width=.8\columnwidth]{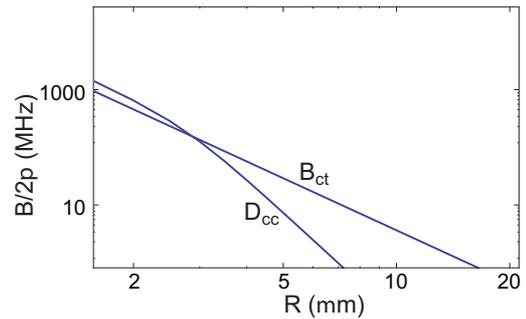}
\par\end{centering}
\caption{\label{fig.sp}The blockade ${\sf B}_{ct}$ and interaction ${\sf D}_{cc}$ strengths for
control atom in  $|s\rangle=|60p_{3/2},m=1/2\rangle$ and target atom in $ |r\rangle=|60s_{1/2},m=1/2\rangle$ states.}
\end{figure}

To proceed we add Eqs. (\ref{eq.Esim1} - \ref{eq.Esim5}) to arrive at a total error
which is  averaged over the lattice. We then  minimize the error with respect to both $\Omega_c$ and $\Omega_t.$ In the averaging procedure ${\sf D}_{cc}$ is only averaged over the $R_{ij}$ between control atom sites and ${\sf B}_{ct}$ is only averaged over $R_{i,j_t}$
where $j_t$ labels the target atom position in the center of the array and $i$ runs over the control atoms.
The $|s\rangle$ and $|r\rangle$ Rydberg levels used for control and target qubits respectively should be chosen to maximize ${\sf B_{ct}}$ and simultaneously minimize ${\sf D}_{cc}.$ This type of asymmetric Rydberg interaction has been invoked previously for multi-bit gates in \cite{Brion2007c,Saffman2009b}. We choose the control-target interaction to be dipole allowed so ${\sf B_{ct}}\sim n^4/R^3$ and seek parameters 
for which the control-control interaction is in the van der Waals limit where ${\sf D_{cc}}\sim n^{12}/R^6$. The asymmetry thus scales as 
${\sf B_{ct}}/{\sf D_{cc}}\sim R^3/n^8$ which suggests using  relatively 
small $n$. Numerical search through different Rydberg levels and lattice periods yielded acceptable results for $|s\rangle=|60p_{3/2},m=1/2\rangle$
and $|r\rangle=|60s_{1/2},m=1/2\rangle$. Interaction curves for these states are given in Fig. \ref{fig.sp}. Note that since we are now using $p$ states the interaction is no longer isotropic and a quantization direction has to be defined. In order to avoid having to perform lattice averaging of the angular dependence we have assumed the quantization axis is perpendicular to the plane of the array so the 
dipole-dipole interaction is isotropic in the plane. This may not be the optimum orientation as far as the gate errors are concerned, so the results obtained need not represent the best possible solution.

Results for the gate errors and corresponding Rabi frequencies are given in Table \ref{tab.Et2}. For the states used  and an optimal lattice period of  about $4.0~\mu\rm m$, the Rydberg interaction strengths are a small fraction of the Rydberg level spacing so interaction induced resonances are not an issue for this approach.  We see that the errors scale slightly slower than linearly with $k$, but they are about three times larger than for the sequential scheme at $k=35$. The errors can be reduced by about 30\% by going to liquid nitrogen temperatures (first line of the table) which increases the Rydberg state radiative lifetimes.  

Although the errors are larger it is worth noting that the number of pulses is about 15 times smaller compared to the sequential approach for $k=35.$ In addition the gate time is $T=3\pi/\Omega_t + 2\pi/\Omega_c =1.1~\mu\rm s$ for $k=35$ compared to about $9~\mu\rm s$ 
for the sequential gate. These differences imply better immunity to technical noise sources, the presence of which will  influence the preferred approach in practice.

\begin{table}[!t]
\centering
\begin{tabular}{|c|c|c|c|c|c|c|c|c|c|}
\hline 
& $d~(\mu\rm  m)$&$k=3$& $8$&$15$&$24$&$35$\\
\hline
error (77 K) &4.2&.02&.03&.07&$.09$&$.16$\\
error (300 K) &4.0&.03&.04&.10&$.14$&$.23$\\
$\Omega_c/2\pi~\rm(MHz)$  &4.0&180&240&300&350&390\\
$\Omega_t/2\pi~\rm(MHz)$  &4.0&0.81&1.6&1.3&1.8&1.6\\

\hline
\end{tabular}
\caption{Simultaneous addressing C$_k$NOT gate errors  averaged over a square lattice. Rydberg states were  $|s\rangle=|60p_{3/2},m=1/2\rangle, |r\rangle=|60s_{1/2},m=1/2\rangle$ with  lifetimes taken  to be\cite{Beterov2009} $\tau_{60p_{3/2}}=(335,148),~\tau_{60s_{1/2}}=(171,97)~\mu\rm s$ at temperatures of (77K,300K). The Rabi frequencies listed are for the 300 K case. }
\label{tab.Et2}
\end{table}

\section{Conclusion}
\label{sec.conclusion}

We have presented two implementations of a C$_k$NOT gate using Rydberg blockade interactions. Using realistic numbers we envision implementation of these gates with $k$ as large as several tens of atoms and errors of only a few percent.  The errors are optimized for $k>15$ at moderate values of $n\sim 75$ for sequential addressing and 
$n\sim 60$ for simultaneous addressing.  The sequential scheme has lower intrinsic errors, while  simultaneous addressing requires fewer laser pulses and is faster, which suggests a lower sensitivity to technical errors.  
These results may be contrasted with the situation for two-qubit gates where higher values of $n$ are necessary for minimum errors at long range. The preference for smaller $n$ for this multi-qubit gate stems from the need to pack as many bits as possible into the strongly interacting resonant dipole-dipole 
regime. This can be done more efficiently at smaller $n$ values. 
 
As we have  shown in \cite{Molmer2011}  a slight modification of these C$_k$NOT gates can be used to efficiently implement Grover search.  
The error analysis for the modified gates is slightly different due to the absence of the target atom in the sequential addressing version. 
For completeness we give the corresponding error expressions in the Appendix. The errors for the modified gates will be comparable to, but slightly less than, those in Table \ref{tab.Et1}. 
Although the C$_k$NOT gates cannot be scaled to arbitrarily large $k$ with acceptable errors the sub-register architecture in \cite{Molmer2011} presents a path to scaling of the search algorithm to arbitrarily large search problems.  The sub-register architecture could in principle also be used to scale our C$_k$NOT gate to larger values of $k$.

The numerical analysis of errors has used Cs atom parameters. We expect similar results would be obtained for Rb, and possibly for other species being considered for Rydberg gate experiments.
The results presented use realistic experimental parameters but could be further optimized along several directions. We have considered a controlled amplitude swap type of operation which requires three Rydberg $\pi$ pulses on the target atom. This could be replaced by two single-qubit 
Hadamards on the target bracketing two Rydberg $\pi$ pulses, which give a C$_k$Z,   to give a C$_k$NOT gate. The Rydberg state related errors on the target qubit would thereby be multiplied by a factor of $2/3$. The Hadamard pulses could also be dispensed with, while keeping only  two Rydberg pulses, using the bright/dark state coupling mechanism from\cite{IRoos2004}. In the simultaneous addressing approach the gate fidelity is limited by the available asymmetry between ${\sf D}_{cc}$ and ${\sf B}_{ct}$. One possible route to improving the performance would be to consider microwave field dressing of the ${\sf D}_{cc}$ F\"orster interaction. Although microwaves have mainly been viewed as a means of strengthening blockade interactions\cite{Bohlouli-Zanjani2007}, in principle it may also be possible to use them to increase the level shift of neighboring states, and thereby reduce the  interaction strength. 
 Furthermore, we anticipate that development of tailored pulse shapes or composite pulses will be a fruitful direction for future improvement of the gate errors in both approaches. 
 
In addition to the use of the modified C$_k$NOT gate for quantum search we expect this gate will be useful for  other quantum algorithms and for quantum error correction. In light of the high cost of implementing the C$_k$NOT using elementary 1- and 2-bit gates our approach may prove advantageous for building multi-qubit processors. 
Analysis of  models for  quantum computation predict error thresholds for fault tolerant operation at 1\%\cite{Knill2005} 
or higher\cite{Wang2011}. For large values of $k$ our predicted errors approach 10\%, and therefore exceed 
known fault tolerance thresholds. However, it should be remembered that the calculated thresholds assume 1- and 2-bit gate operations, many of which would be needed to implement a single C$_k$NOT operation. It therefore remains 
an open question as to the usefulness of many-bit gates with moderate error levels for fault tolerant implementations of quantum algorithms.

\begin{acknowledgements}
This research was supported by the EU integrated
project AQUTE, the IARPA MQCO program,  DARPA, and  NSF award
PHY-0969883.
\end{acknowledgements}

\bibliographystyle{spphys}       


\begin{thebibliography}{10}
\bibitem{RydbergBadLuck}
It is also necessary to ensure no accidental resonances where the Rydberg laser
  frequencies $\pm \omega_{10}$ or $\pm (\omega_{10}\pm {\sf B})$ point to a
  different Rydberg level. We assume that the Rydberg level and lattice spacing
  $d$ can be suitably chosen to avoid such problems. As a check on the
  sensitivity of the gate errors to $\omega_{10}$ we have recalculated Table
  \ref{tab.Et1} using a value of $\omega_{10}$ that is artificially reduced by
  a factor of 10. This gives up to 50\% error increase for $k=3$ where the
  errors are in any case very small, and $<10\%$ increase for $k\ge 8.$
\end{thebibliography}

%
%


\section{Appendix}

Here we give the error estimates  for the sequentially addressed modified C$_k$NOT operations used in \cite{Molmer2011}. The analysis follows that in  Sec. 
\ref{sec.errors}
with the only difference being that  we apply the $2k$ $\pi$ pulses to the ``control" atoms, but there is no target atom. 
In this way a conditional phase is picked up on the qubit register which is used to implement the Grover iterations. The error estimates are therefore slightly smaller and are given by

\begin{eqnarray}
E_{se,c,1}&=&\frac{\pi}{\Omega\tau} \left( 2k-3+\frac{3}{2^k}\right)\nonumber\\
E_{se,c,2}&=&\frac{\pi\Omega}{4{\sf B}^{2}\tau}\left(k^{2}-4k+6-\frac{6}{2^{k}}\right)\nonumber\\
 E_{r,c,1}&=&\frac{\Omega^{2}}{2{\sf B}^{2}}\left(k-2+\frac{1}{2^{k-1}}\right)\nonumber\\
E_{r,c,2}&=&\frac{\Omega^{2}}{\omega_{10}^{2}}\left( 1-\frac{1}{2^{k}}\right) +\frac{\Omega^{2}}{\left(\omega_{10}\pm {\sf B}\right)^{2}}\left(\frac{k}{2}+\frac{1}{2^{k}}-1\right).\nonumber
\end{eqnarray}

The combined gate error is 
\begin{eqnarray}
E&=&\frac{\pi\Omega}{4{\sf B}^2\tau}\left[k^2-4k+6\left(1-\frac{1}{2^{k}}\right) \right]\nonumber\\
&&
+\frac{ 2\pi  }{\Omega\tau}\left(k-\frac{3}{2}+\frac{3}{2^{k+1}}\right) \nonumber\\
&&
+\frac{ \Omega^2}{2 {\sf B}^2}\left(k -2+\frac{1}{2^{k-1}}\right)\nonumber\\
&&
+\frac{ \Omega^2}{2 ({\sf B}\pm \omega_{10})^2}k\nonumber .
\label{eq.EtG}
\end{eqnarray}
In the limit of ${\sf B}\tau\gg k\gg 1 $ the optimum Rabi 
frequency $\Omega_{\rm opt}$  and the minimum gate error are 
the same as in Eqs. (\ref{eq.Omegaopt},\ref{eq.Eopt}).
There is no difference since in the limit of $k\gg 1$ the addition of the three target atom pulses in the C$_k$NOT protocol compared to a Grover iteration step has negligible impact on the error.

\end{document}